\documentclass[aps, prx, reprint,10pt,longbibliography,superscriptaddress,nofootinbib]{revtex4-2}

\usepackage[caption=false]{subfig}
\usepackage{amsmath, amssymb, amsfonts, amsthm, bm, mathrsfs, bbm, enumerate,mathtools}
\usepackage{graphicx}
\usepackage{epstopdf}
\usepackage{color}
\usepackage[usenames,dvipsnames]{xcolor}
\usepackage[citecolor=blue, colorlinks=true, urlcolor=blue, linkcolor=blue]{hyperref}
\usepackage{bbold}
\usepackage{feynmp-auto} 

\usepackage{soul}
\usepackage{float}

\newcommand\varpm{\mathbin{\vcenter{\hbox{%
  \oalign{\hfil$\scriptstyle+$\hfil\cr
          \noalign{\kern-.3ex}
          $\scriptscriptstyle({-})$\cr}%
}}}}

\begin{document}

\title{Chiral Pseudo Spin Liquids in Moir\'e Heterostructures}
\author{Clemens Kuhlenkamp}
\affiliation{Department of Physics, Technical University of Munich, 85748 Garching, Germany}
\affiliation{Munich Center for Quantum Science and Technology (MCQST), Schellingstr. 4, D-80799 M{\"u}nchen, Germany}
\affiliation{Institute for Quantum Electronics, ETH Z\"urich, CH-8093 Z\"urich, Switzerland}
\author{Wilhelm Kadow}
\affiliation{Department of Physics, Technical University of Munich, 85748 Garching, Germany}
\affiliation{Munich Center for Quantum Science and Technology (MCQST), Schellingstr. 4, D-80799 M{\"u}nchen, Germany}
\author{Ata\c{c} Imamo\u{g}lu}
\affiliation{Institute for Quantum Electronics, ETH Z\"urich, CH-8093 Z\"urich, Switzerland}
\author{Michael Knap}
\affiliation{Department of Physics, Technical University of Munich, 85748 Garching, Germany}
\affiliation{Munich Center for Quantum Science and Technology (MCQST), Schellingstr. 4, D-80799 M{\"u}nchen, Germany}
\begin{abstract}

We propose multi-layer moir\'e structures in strong external magnetic fields as a novel platform for realizing highly-tunable, frustrated Hubbard physics with topological order. Identifying the layer degree of freedom as a pseudo spin, allows us to retain SU(2) symmetry while controlling ring-exchange processes and concurrently quenching the kinetic energy by large external magnetic fields. This way, a broad class of interacting Hubbard-Hofstadter states and their transitions can be studied. Remarkably, in the limit of strong interactions the system becomes Mott insulating and we find chiral pseudo spin liquid phases which are induced by the magnetic field. We find that this topologically ordered state remains exceptionally stable towards relevant perturbations. We discuss how layer pseudo-spin can be probed in near-term experiments. As the magnetic flux can be easily tuned in moir\'e systems, our approach provides a promising route towards the experimental realization and control of topologically ordered phases of matter.

\end{abstract}

\date{\today}

\maketitle

\section{Introduction}

Frustrated electronic systems are expected to exhibit a variety of exotic quantum phases, including fractional quantum Hall states and spin liquids, which host intrinsic topological order and emergent gauge fields~\cite{Wen2007, Savary2016, Knolle2019}. While these systems have attracted tremendous theoretical and experimental interest, the decade long quest for the experimental realization of such long-range entangled states has been challenged by their fragility. In particular, when topological order is encoded in the spin degree of freedom, a characterization and manipulation of the excitations becomes extremely challenging. Even in theoretical models, spin liquids often appear only for very specific parameters, which makes finding good candidate materials difficult~\cite{Knolle2019,Kee2016}. Consequently, conclusive evidence for spin liquids in solids is hard to find, which makes it essential to identify tunable solid-state platforms that allow for novel probes of the spin liquid states.

The kinetic energy of electrons can be controllably quenched by large, external magnetic fields in moir\'e heterostructures of transition metal dichalcogenides (TMDs). While these magnetic fields eventually polarize the electron spin, we can retain an $SU(2)$ symmetry by considering synthetic layer pseudo-spins instead of electronic spins~\cite{Eisenstein1992,Kim2017,schwartz21}. The lowest bands of such systems are described by effective triangular lattice Hubbard models, which makes them particularly promising settings to study the interplay of geometric frustration and strong interactions~\cite{regan_20,yuya_20,tang_20,Kim2022,Vishwanath2021}. Here, we show that the magnetic field induces a particularly rich phase diagram, ranging from Hofstadter physics for small interactions to spin-liquid phases in the Mott insulating regime~\cite{hofstadter76,Anderson1973,Laughlin87,Wen89,Motrunich2006,Vishwanath2021}. In our setting, the magnetic field reliably tunes ring-exchange processes, which stabilize exceptionally robust chiral spin liquids (CSL)~\cite{Motrunich2006,Sheng2017}. Due to the large moir\'e unit-cells it is possible to insert high enough flux to explore the full spectrum of the Hofstadter butterfly~\cite{hofstadter76,Geim2013}. This enables us to study a large variety of previously inaccessible phases, including exotic insulator-to-insulator transitions between topological charge and spin sectors.

Our work is structured as follows. We introduce the effective frustrated Hubbard model for twisted TMD bilayer structures in the presence of a magnetic field in Sec.~\ref{sec:model} and discuss details of the model in App.~\ref{App:A}. The stabilized phases are presented in Sec.~\ref{sec:phases}. We analyze the robustness of topologically ordered spin liquid phases in Sec.~\ref{Sec:robust}. Experimental signatures of the layer pseudo spin liquid are proposed in Sec.~\ref{sec:signatures}. We provide a discussion and an outlook in Sec.~\ref{sec:outlook}.

\begin{figure*}[t!]
\includegraphics[width=0.95\textwidth]{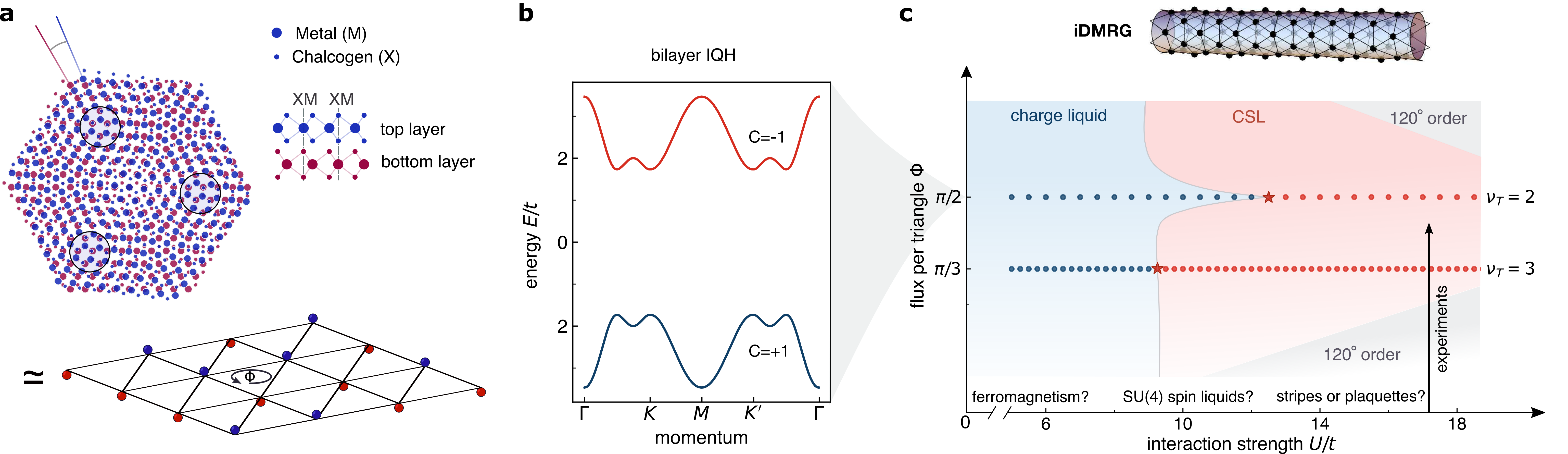}
\caption{\textbf{Setup and schematic phase diagram.} \textbf{a)} Twisted AB-stacked homobilayer TMD setup to realize a triangular Hubbard model. Top and bottom layers are drawn in blue and red. Electrons with pseudo-spin up (down) in the effective Hubbard model are depicted as blue (red) spheres in the bottom panel. \textbf{b)} Typical topological band structure of the Hofstadter model at $\pi/2$ flux and vanishing interactions. \textbf{c)} Schematic phase diagram of the half-filled triangular Hubbard model as a function of flux and interactions. Circles represent iDMRG data obtained on cylinders of circumference $L_y=6$ and stars mark phase transitions. Remarkably, for intermediate to large interactions the dominant fraction of the phase diagram is a Chiral Spin Liquid (CSL). For weak interactions, the magnetic flux gives rise to a large variety of Hofstadter states, which can directly transition to the CSL. In the limit $\Phi\rightarrow 0$, the system recovers an approximate SU(4) symmetry once the intrinsic electronic spin is no longer polarized.}
\label{fig:1}
\end{figure*}

\section{Frustrated Hubbard physics in twisted TMD bilayers\label{sec:model}}
Heterostructures of two dimensional (2D) materials form moir\'e patterns, either when there is a lattice constant mismatch between the materials, or when some of the layers are twisted against each other. This commonly results in honeycomb and triangular structures, which imprint lattices on the electrons. Motivated by recent work~\cite{Vishwanath2021, Yao2021}, we consider triangular moir\'e patterns, in which two lattices in different layers coincide spatially. This can be achieved in multiple ways: in 'sandwich' stacked trilayer systems; for AB-stacked homobilayers with a twist~\cite{Vishwanath2021,Xu2022}, as shown in Fig.~\ref{fig:1} a); via twisted hexagonal boron nitride structures in proximity to a homobilayer TMD~\cite{Yao2021,Woods2021}; or via electrostatically imprinted potentials~\cite{Huber2022}. Depending on the depth of the lattice potential~\cite{Hu2021}, the lowest lying moir\'e band of the TMD heterostructure can then realize an effective triangular Hubbard model, with two electronic degrees of freedom: spin and layer. Here we explore the phase diagram of these models by tuning the external magnetic field $B_z$. For $B_z \rightarrow 0$ the system can be described by a generalized Hubbard model with an approximate SU(4) symmetry~\cite{Vishwanath2021}. In the strong magnetic field limit, the Zeeman effect will fully polarize the electrons. Consequently their electronic spin can be discarded and the TMD system is described by an interacting Hofstadter-Hubbard model on the triangular lattice with two flavors:
\begin{equation}
\begin{aligned}
\hat{H} &= - t \sum_{\mathclap{\langle ij\rangle, \sigma = \lbrace T,B \rbrace}} e^{i \phi_{ij}} \; c^\dagger_{i,\sigma} c_{j,\sigma} + \mathrm{h.c.} + t_\perp \sum_{i} c^\dagger_{i,T} c_{i,B}+ \mathrm{h.c.} \\
&+ U \sum_i n_{i,T} \; n_{i,B},
\end{aligned}
\label{Eq:hamilt_fermions}
\end{equation}
where we have assumed that longer-range interactions have been screened by nearby gates. We discuss the details of the effective model and subleading contributions in App.~\ref{App:A}. The operator $c^\dagger$ ($c$) creates (annihilates) spin polarized fermions and the index $\sigma \in \lbrace T,B \rbrace$ labels the top and bottom layer of the heterostructure, see Fig.~\ref{fig:1} a) for an illustration. Electrons are subject to a hopping term with strength $t$ and an inter-layer tunnel coupling $t_\perp$. In AB stacking, one of the layers is rotated by $60^\circ$ which exchanges the $K$ and $K'$ points. Due to spin-valley locking, this strongly suppresses $t_\perp$. For sandwich structures and twisted h-BN imprinted potentials, $t_\perp$ can be made vanishingly small via an insulating barrier. In the following we analyze Eq.~\ref{Eq:hamilt_fermions} and assume $t_\perp=0$ throughout. We interpret electrons in the top (bottom) layer as having pseudo spin $ +1/2$ ($-1/2$). The magnetic field does not couple directly to the pseudo-spin, and its only effect is the generation of Peierls phases $\phi_{ij}$. These time-reversal symmetry breaking phases are crucial for the following discussion and constitute the main difference compared to previously proposed setups~\cite{tang_20,Kim2022}.

Hamiltonian~\ref{Eq:hamilt_fermions} then describes lattice versions of quantum Hall systems, which give rise to a large variety of phases. In the absence of interactions the system realizes a Hofstadter model that hosts a multitude of topologically non-trivial electronic bands which are induced by the magnetic field; see Fig.~\ref{fig:1} b) for an example. As solid state systems are generically interacting, it is crucial to study the fate of these bands once electron repulsion is considered. While some states directly connect to the Hofstadter model at $U=0$, interactions stabilize exotic phases of matter. One such example are excitonic superfluids, which arise when pseudo-spin symmetry is broken spontaneously by the interactions. Another, even more exciting possibility, is the formation of states with intrinsic topological order, such as fractional Chern insulators. As opposed to integer Chern insulators they feature emergent gauge fields, anyonic excitations and long-range entanglement~\cite{Wen90,Wen2006}.

As correlated insulating phases give rise to particularly interesting spin physics, we consider a single electron per moir\'e unit cell $n_e = 1$ from now on. This allows the system to become Mott insulating for strong enough interactions, which is the natural regime for TMDs~\cite{Wang21,Park2021,Smolenski2021}. This limit is understood by eliminating the doubly occupied subspace with a Schrieffer-Wolff transformation~\cite{Motrunich2006}, leading to the following effective Hamiltonian
\begin{equation}
\begin{aligned}
    \hat{H}_{\mathrm{eff.}} &= \frac{1}{U} \sum_{ij,\sigma \sigma'} t_{ij}t_{ji}\; (c_{i\sigma}^\dagger c_{i\sigma'}) \;  (c_{j\sigma'}^\dagger c_{j\sigma}) \\
    -& \frac{2}{U^2}\sum t_{ij} t_{jk} t_{ki} \; (c_{i\sigma}^\dagger c_{i\beta}) \, (c_{k\beta}^\dagger c_{k\alpha}) \, (c_{j\alpha}^\dagger c_{j\sigma})  \\
    &+ \mathcal{O}(t^4/U^3),
    \label{Eq:hamilt_fermionrep}
\end{aligned}
\end{equation}
which is equivalent to an effective spin system
\begin{equation}
\begin{aligned}
    \hat{H}_{\mathrm{eff.}} =& J_H\sum_{\langle i j \rangle} \mathbf{S}_i \cdot \mathbf{S}_j + J_\chi \sum \mathbf{S}_i \cdot( \mathbf{S}_j \times \mathbf{S}_k)\\
    &+ \mathcal{O}(t^4/U^3) ,
    \label{Eq:hamilt_spins}
\end{aligned}
\end{equation}
with anti-ferromagnetic Heisenberg interactions $J_H$ and a chiral spin coupling $J_\chi$. The couplings are related to the Hofstadter-Hubbard model parameters as follows
\begin{equation}
    J_H = 4 t^2/U, \quad J_\chi = 24 \,\sin(\Phi)\; t^3/U^2,
\end{equation}
where $\Phi$ is the magnetic flux per triangle, which is given by $\phi_{ij} + \phi_{jk} + \phi_{ki}$ if $i,j,k$ label sites on a single triangle. 

Although the triangular lattice is frustrated, the ground state of the pure Heisenberg model, obtained for $U=\infty$, is a co-planar 120 degree (pseudo-)magnet~\cite{Miyashita84,HuseElser}. However, when interactions are lowered or the magnetic flux is increased, the ground state of the fermion system is still a subject of debate. We will show that $J_\chi$ can melt the 120 degree order and give way to an exceptionally stable CSL. The hallmark of this exotic phase is a fractionalization of the spins into spinons $f_\alpha$, which is captured by a parton ansatz $\vec{S} =\frac{1}{2} \sum_{\alpha,\beta=0}^1 f_\alpha^\dagger \vec{\sigma}_{\alpha\beta} f_\beta$. Expressing Eq.~\ref{Eq:hamilt_spins} in terms of the spinons leads to a Hamiltonian, identical to Eq.~\ref{Eq:hamilt_fermionrep} with electrons replaced by the spinons. The price to pay in this representation is a single occupancy constraint $\sum_\alpha f_{i,\alpha}^\dagger f_{i,\alpha} = 1$ which must be imposed on each site $i$. This is similar to the Schrieffer-Wolff construction in Eq.~\ref{Eq:hamilt_fermionrep}, where we projected out double occupancies to describe the Mott insulator. Spinons evolving with Eq.~\ref{Eq:hamilt_fermionrep} may move freely in a correlated fashion if self-consistently generated hopping terms acquire non zero expectation values $\sum_\beta \langle f_{i,\beta} f^\dagger_{j,\beta}\rangle  \neq 0$, but remain confined for conventional magnetic phases such as the 120 degree order. A mean-field decoupling around such configurations leads to simple trial Hamiltonians for the spin liquid:
\begin{equation}
H_{\mathrm{trial}} = \sum_{\langle ij \rangle, \alpha} \tilde{t}_{ij} f_{i\alpha}^\dagger f_{j\alpha}.
\label{Eq:trial_hamilt}
\end{equation}
This illustrates one of its key properties: a stable mean field solution for the chiral spin liquid exists, when the (self-consistently determined) hopping matrix $\tilde{t}_{ij}\sim \sum_\beta \langle f_{i,\beta} f^\dagger_{j,\beta}\rangle  $ breaks time reversal symmetry and induces topological Chern bands. The ground state of the CSL can be thought of as a Chern insulator composed of spinons. Projecting out double occupancies to satisfy the single occupancy constraint then gives the spinons anyonic character~\cite{Wen2007,Motrunich2006}. Despite the fragility of the CSL at zero flux and the fact that a Mott state retains a nonzero (but small) fraction of double occupancies, we will show that magnetic fields strongly favor the formation of a quantum spin liquid, see Fig.~\ref{fig:1} c) for a schematic phase diagram. 

As a topologically ordered state, the CSL is robust to any small perturbation. We study the stability of the CSL to the most relevant subleading contributions to our model, including longer-range interactions, layer SU(2) symmetry breaking, and intersite ferromagnetic exchange interactions; see Sec.~\ref{Sec:robust}. For realistic estimates of these subleading contributions, we find that the CSL remains remarkably stable. This is a consequence of the explicit time-reversal symmetry breaking by the external magnetic field.
 
While this discussion assumed strong magnetic fields to fully polarize the electronic spin, we point out that rich physics also arises when Zeeman splittings are small. Then, the system has additional approximate symmetries which have been proposed to give rise to more exotic SU(4) spin liquids, while the 120 degree state is replaced by stripe order, plaquette order or SU(4) symmetry broken trimer states~\cite{Vishwanath2021,Keselman23}. This regime is highlighted in the bottom of Fig.~\ref{fig:1}c). How this and other phases interplay with the Hofstadter states is an interesting open question, which is beyond the scope of this work.

Furthermore our discussion based on the effective spin model neglects higher order ring-exchange processes, which become increasingly important close to the Mott transition. In order to fully capture the properties of the system, we instead study the electronic Hamiltonian of Eq.~\ref{Eq:hamilt_fermions} directly, without projecting out the higher energy subspace~\citep{Moore2020}. In the following, we emphasize the rich physics of the model by studying a subset of phase transitions from Hofstadter states to the CSL.

\section{Phases of the pseudo-spin Hofstadter-Hubbard model \label{sec:phases}}
We consider in detail the half-filled triangular Hubbard model  deep inside the Hofstadter butterfly regime with two fluxes $\Phi \in \lbrace \pi/3,\pi/2\rbrace$. These flux values  correspond to total filling factors of $\nu_T = 2 \pi n_e/2 \Phi \in \lbrace 3, 2 \rbrace$, where $2\Phi$ is the flux per unit cell and $n_e$ is the number of electrons per unit cell ($n_e=1$ at half-filling, which is what we consider throughout this work). 
For these flux values various exotic phases of matter can be stabilized. For example, at weak interactions the system could realize gapped bilayer integer quantum Hall (IQH) and gapless excitonic insulators, respectively. For large interactions, these phases compete with magnetic ordering and exotic spin liquids. Given the insights from the effective spin model in Eq.~\ref{Eq:hamilt_spins} the precise value of the external flux only determines the ratio of $J_\chi/J_H$ in the Mott insulator. However, in the quantum Hall regime its value is essential, as it determines charge gap. In order to determine which phases are stabilized by the microscopic Hofstadter-Hubbard model, we resort to numerical methods.

\textbf{Method.---}Studying fermions in magnetic fields poses a significant numerical challenge. Here, we use matrix product states (MPS) to obtain an unbiased variational approximation of the many-body wave function. This Ansatz allows for an expansion in terms of the entanglement, that is controlled by the maximal bond dimension $\chi$ of the MPS. If both the quantum Hall states and the CSL are gapped, they can be efficiently represented in terms of an MPS on a cylindrical lattice, that is finite in one direction but infinite in the other, because the total amount of entanglement is finite for such lattice geometries. This method has been successfully applied to phases with intrinsic topological order, which lead to a better understanding of fractional quantum Hall (FQH) and fractional Chern insulators; see e.g.~\cite{Rezayi2015,Grushin2015,Haldane2022}. We variationally optimize the MPS by infinite Density Matrix Renormalization Group (iDMRG), implemented via the TeNPy library \cite{Hauschild2018}. Since the bond dimension, and hence the numerical costs, grow exponentially with the cylinder circumference, in this work we focus on $L_y = 6$, which fits both fluxes $\Phi=\pi/2, \pi/3$ with periodic boundary conditions.

Working in the infinite limit along the x-direction, we can directly obtain the correlation functions as well as the correlation length from the transfer matrix of the corresponding MPS unit cell. We use these methods to determine the ground-state phase diagram of the Hofstadter-Hubbard model given by Eq.~\ref{Eq:hamilt_fermions}. Even though the precise points of the phase transitions may shift depending on the cylinder circumference, we expect the structure of the phase diagram to be similar in the two dimensional limit \cite{Moore2020}. To be able to perform the simulations with high bond dimensions and therefore small truncation errors, we utilize the $\mathrm{U}(1)\times \mathrm{U}(1)$ symmetries generated by the z-component of pseudo spin $\sum_i S^z_i = \sum_i \hat{n}^T_i- \hat{n}^B_i$ and particle conservation $\sum_i \hat{n}^T_i + \hat{n}^B_i$ as well as translation symmetry along the y-direction \cite{Motruk2016, Ehlers2017}, which determine the $S_z$, $Q$, and $k_y$ quantum numbers, respectively.

\begin{figure}
\includegraphics[width=0.95\columnwidth]{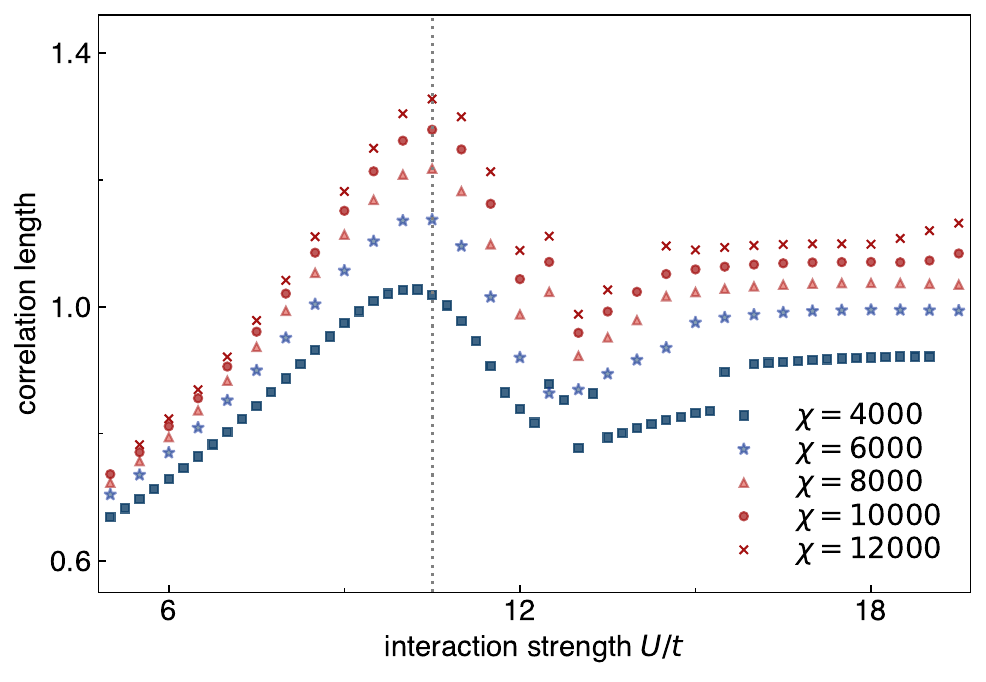}
\caption{\textbf{Ground state correlation length for $\Phi=\pi/2$.} We show the correlation length in the charge sector $(S_z,Q,k_y)=(0,0,0)$ as a function of $U/t$ for different values of the bond dimension $\chi$. Around $U\simeq 10.5t$ the correlation length grows with bond dimension, which is indicative of phase transition melting the bilayer quantum Hall state. The dotted grey line serves as a guide to the eye.}
\label{fig:2}
\end{figure}

\textbf{Quantum phase transition at $\mathbf{\Phi = \pi/2}$.---}
\begin{figure*}[!ht]
\includegraphics[width=0.95\textwidth]{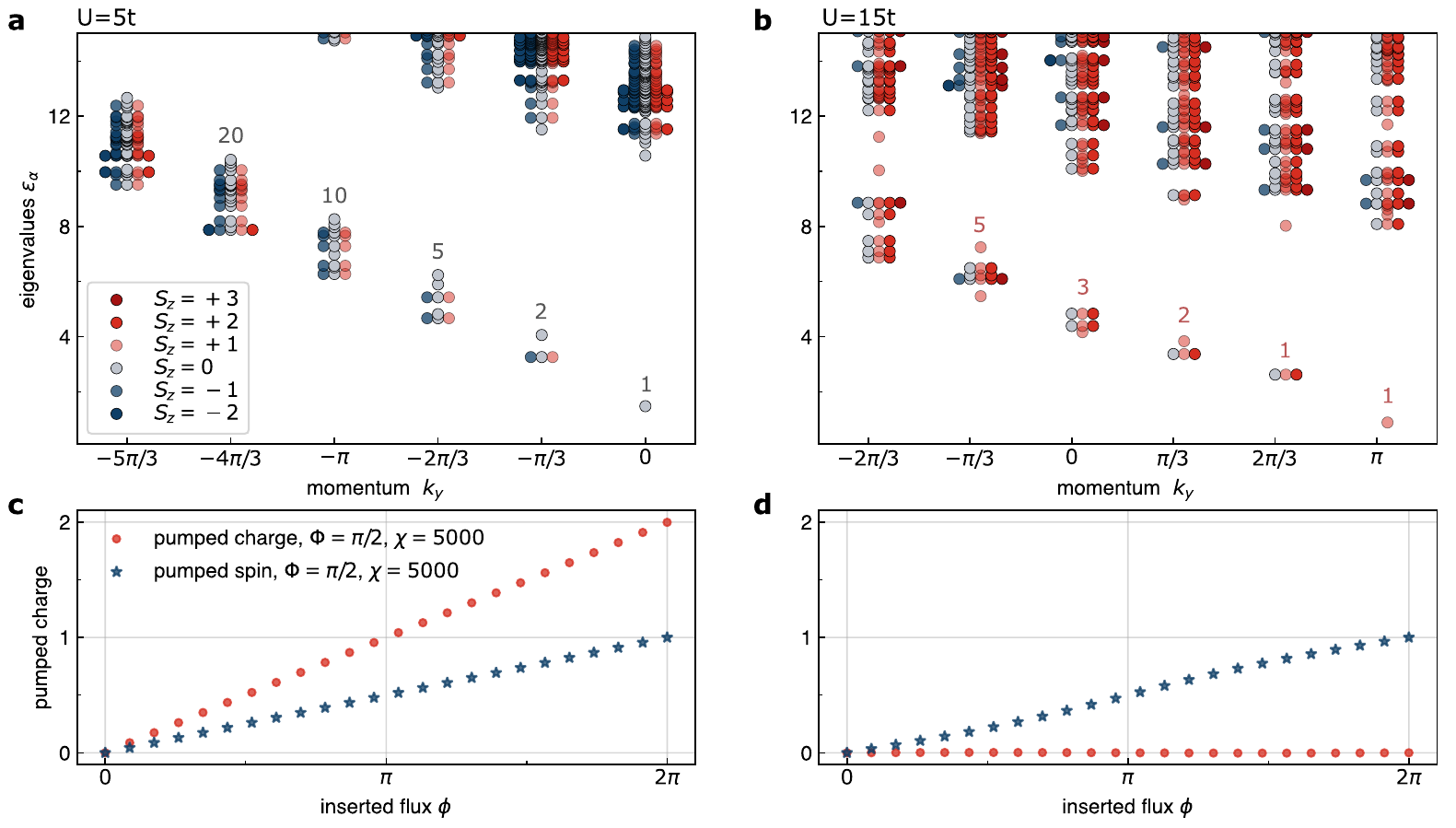}
\caption{\textbf{Topological properties for $\Phi=\pi/2$.} Edge properties and charge pumping for $\Phi=\pi/2$ for $U=5t$ \textbf{(a,c)} and $U=15t$ \textbf{(b,d)}. \textbf{a)} Ground state entanglement spectrum in the IQH phase for $\chi=10000$ and charge $Q=0$. The spectrum obeys a (1,2,5,10,\dots) counting rule characteristic for an edge theory of two chiral bosons, which suggests an IQH state in both the top and bottom layer. \textbf{b)} Ground state entanglement spectrum in the CSL for $\chi=10000$ and charge $Q=0$. The spectrum obeys the (1,1,2,3,5,\dots) counting rule for the edge theory of the CSL phase. \textbf{(c,d)} Pumped charge under flux insertion through the cylinder. Red and blue dots indicate flux that is inserted equally and oppositely in the two layers. While the pumping in \textbf{c)} is a direct consequence of two charge-carrying edge modes of the bilayer IQH state, the charge pumping in the CSL regime \textbf{d}) vanishes entirely, consistent with the separation of spin and charge degrees of freedom while the spin remains quantized.}
\label{fig:3}
\end{figure*}
For $\Phi=\pi/2$ flux per triangle and small $U/t$ the system is in a bilayer IQH state, defined by a fully filled topological band with Chern number $C=1$ for each pseudo spin, see Fig.~\ref{fig:1}b). We find signatures of a phase transition by studying the ground state correlation length as a function of interactions $U/t$, which is shown for operators carrying the quantum numbers $(S_z,Q,k_y)=(0,0,0)$ in Fig.~\ref{fig:2}. As interactions are increased, the correlation length grows significantly with bond dimension around a critical interaction strength of $U_c \sim 10.5 t$, which indicates a gap closing phase transition. In contrast to the bilayer IQH state, identifying the phases for $U>U_c$ is more subtle. The most relevant competing states are 120 degree spin-order, tetrahedral spin-order, excitonic insulators, and the chiral spin liquid.

We shed light on the large $U$ phase by noticing that the enhanced correlation length is accompanied by a simultaneous reorganization of the half-cylinder entanglement spectrum, shown in Fig.~\ref{fig:3}~(a,b). The entanglement spectrum directly encodes the energy levels of the edge theory on a half-infinite cylinder, which are distinct for the bilayer IQH and the other candidate phases. The edge theory of the bilayer IQH for small $U/t$ is given by two chiral modes, one for each layer. Their excitations are understood as follows: For $k_y=0$ there is a unique state where neither of the edge modes is excited, leading to a single dominant entanglement eigenvalue. To create a momentum $k_y = 1\cdot 2\pi /L_y$ excitation, one can shift the lowest lying electron by one momentum quanta, and promote it to the state just above the Fermi level. As this can be done in both layers, we find two such excitations, leading to two entanglement eigenvalues. In the appropriate basis these excitations decouple completely and correspond to total-density and spin-density wave excitations. Continuing this counting, one finds that the momentum resolved entanglement spectrum for increasing momenta $k_y$ is given by ($1,2,5,10,\dots$) entanglement eigenvalues in each spin sector, see Fig.~\ref{fig:3}~a). These density- and spin-wave modes are arranged in a representation of an underlying $U(1)\otimes SU(2)_1$ algebra, as expected for the boundary of a double copy of an IQH state. For $U > U_c$, on the other hand, the ground state loses several edge excitations; see Fig.~\ref{fig:3}~b). This is natural once the system turns Mott insulating, at which point density waves acquire a finite energy cost. The spectrum then consists only of spin waves described by a representation of the $SU(2)_1$ algebra, which leads to a ($1,1,2,3,5,\dots$) counting. The edge is therefore captured by a chiral $SU(2)_1$ Wess-Zumino-Witten model, which matches the edge theory of the chiral spin liquid~\cite{Wen1991a, Moore1997}.

We further analyze the topological character of the bilayer IQH and the suspected CSL by inserting fluxes through our cylinder, which realizes Laughlins charge pump~\cite{Laughlin1981,Zaletel2013}. The IQH state responds to the insertion of $2\pi$ magnetic flux by transferring a single charge from one end of the cylinder to the other in each layer, illustrated in Fig~\ref{fig:3}~c) (red markers). To couple to the charge-neutral spin edge modes, one has to thread opposite fluxes in the two layers. The IQH state then pumps one electron in the top and one hole in the bottom layer each carrying a spin of $1/2$, leading to a pseudo-spin transfer of unity, see Fig~\ref{fig:3}~c) (blue markers). Combining these results, we infer that the excitations carry both charge and spin.

For $U>U_c$ the system no longer pumps electric charge when threading magnetic fluxes, consistent with a Mott insulating state. However, it still exhibits quantized spin transfer for oppositely inserted fluxes, see panel d) of Fig.~\ref{fig:3}. Inserting $\pi$ flux for the electrons corresponds to a $2\pi$ flux insertion in the effective spin Hamiltonian Eq.~\ref{Eq:hamilt_spins}. 
From this point of view we pump a single spin after threading $4\pi$ spin-flux, realizing a fractional spin Hall effect. This can be intuitively understood via the Kalmeyer-Laughlin construction, where the spin system is mapped to a half-filled Bose-Hubbard model in the presence of a fictitious background magnetic field~\cite{Laughlin87}. Our observed spin pumping is then explained by the formation of a $\nu=1/2$ bosonic FQH state which is identified as the chiral spin liquid in the spin picture. As such, excitations of the CSL are semions, which are abelian anyons with a statistical phase $\theta=\pi/2$. This leads to a doubly degenerate ground state on cylinders in the thermodynamic limit $L_y =\infty$, which reside in the $k_y =0$ and $k_y=\pi$ momentum sectors respectively. In our geometry, the ground state energies in these sectors approach each other and cross shortly after the transition, which is a finite size effect of our simulations~\cite{Moore2020,Sheng2017}. The same conclusion can be reached starting from a purely fermionic model, although the discussion is more involved~\cite{Wen2007}. The fractional spin-Hall effect and the absence of charge transport therefore serves as a direct signature of a fractional Hall state for spin. Putting these findings together we identify the phase into which the bilayer IQH phase transitions when increasing interactions to $U>U_c$ as a CSL. 

\textbf{Competition between Hall states and spin liquids at $\mathbf{\Phi = \pi/3}$.---}
\begin{figure}[!ht]
\includegraphics[width=1\columnwidth]{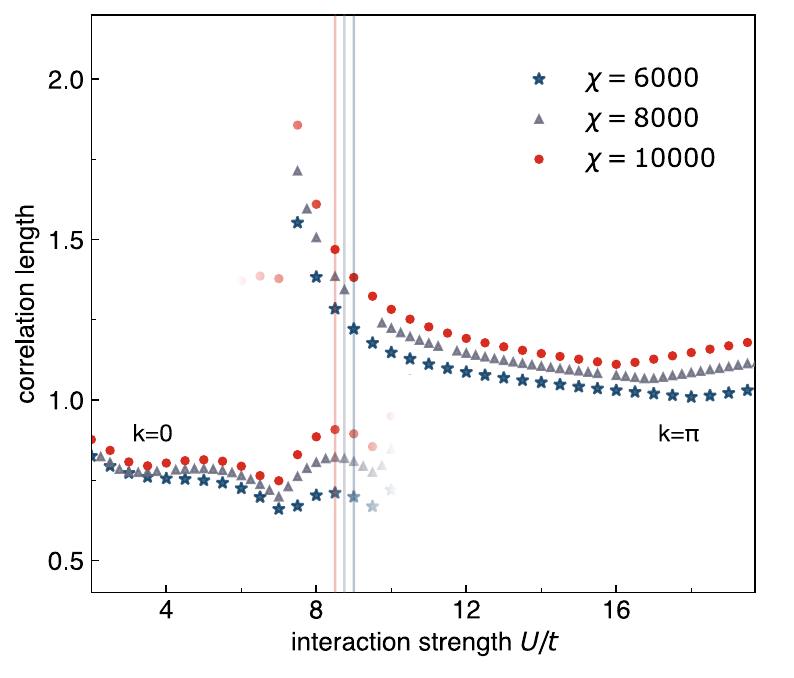}
\caption{\textbf{Transition for $\Phi = \pi/3$.} The ground state correlation length for operators in the sector ($S_z$,$Q$,$k_y$) = (0,0,0). In the lowest energy $k_y=\pi$ state, the correlation length strongly increases around $U\simeq 8.5t$, indicating a sharp transition to a chiral spin liquid. The energy crossing of the $k=0$ and $k=\pi$ states is indicated by the vertical lines and is a finite size effect. The two momentum states are expected to be the degenerate ground states of the CSL in the thermodynamic limit.}
\label{fig:4}
\end{figure}
For $\nu_T=3$, we no longer expect a quantized Hall conductance as the Landau bands are partially filled. This opens the possibilities for other charge liquids, such as excitonic superfluids. Here, we choose unit cells of size $L_x=3$, which allows us to reliably prepare states with fixed momenta along the y-direction; see supplemental materials~\cite{supp}. 

We find that for small $U/t$ the system does not exhibit any quantized pumping; see supplementary materials~\cite{supp}. Although it is suggested by the continuum limit, we do not find long range ferromagnetic correlations of the Hall liquid, which may be a feature of the Hofstadter-Hubbard model at half filling. While we cannot uniquely identify the nature of this state, its response is consistent with a featureless Hall state~\cite{Eisenstein2014}. However, once interactions are raised a sharp transition occurs around $U\simeq 8.5t$, as evidenced by the growing correlation length; see Fig.~\ref{fig:4}.

We find that the transition (on finite cylinders) occurs in steps: First, for small $U/t$ the ground state is found at zero momentum. As $U$ increases the correlation length in the $k_y =\pi$ sector diverges. Then, quickly after this divergence the $k_y =\pi$ state becomes the new ground state of the system. Afterwards the energy splitting between the two states remains roughly constant~\cite{supp}. 

There are several possible candidate states once the system becomes insulating. Following the analysis of the previous section we can identify the phase for $U\geq 8.5t$ as a CSL~\cite{supp}. In particular, the state exhibits a fractional spin-Hall effect and shows the characteristic half-cylinder entanglement spectra. Although the $k_y=0$ state is higher in energy than the $k_y=\pi$ state for $U\geq 8.5t$, the edge theories of both states are in good agreement with the $SU(2)_1$ WZW model describing the boundary of the CSL.

While the robust spin liquid is a universal feature of the large $U$ phase for both fluxes, the charge to spin liquid transitions are clearly distinct for $\nu_T=2$ and $\nu_T=3$. Most notably for $\nu_T=3$ the CSL appears already for much weaker interactions compared to $\nu_T=2$. This indicates that the charge liquid at $\nu_T=3$ is less stable, than the $\nu_T=2$ bilayer IQH state. A sketch of the expected phase boundaries is shown in Fig.~\ref{fig:1} c). This behaviour is reminiscent of the competition between Wigner crystals and Hall states in electron gases, where integer and fractional quantum Hall states extend further into the gapped crystalline phase than their gapless counterparts~\cite{Shayegan21,Smolenski2021}.

\section{Robustness of the CSL to perturbations}
\label{Sec:robust}

Gapped topological phases are roubst toward perturbations. Here, we characterize the robustness of the CSL to subleading corrections to the effective Hamiltonian~\eqref{Eq:hamilt_fermions}. 

\textbf{Nearest-neighbor interactions.---}
\begin{figure*}[t]
\includegraphics[width=0.95\textwidth]{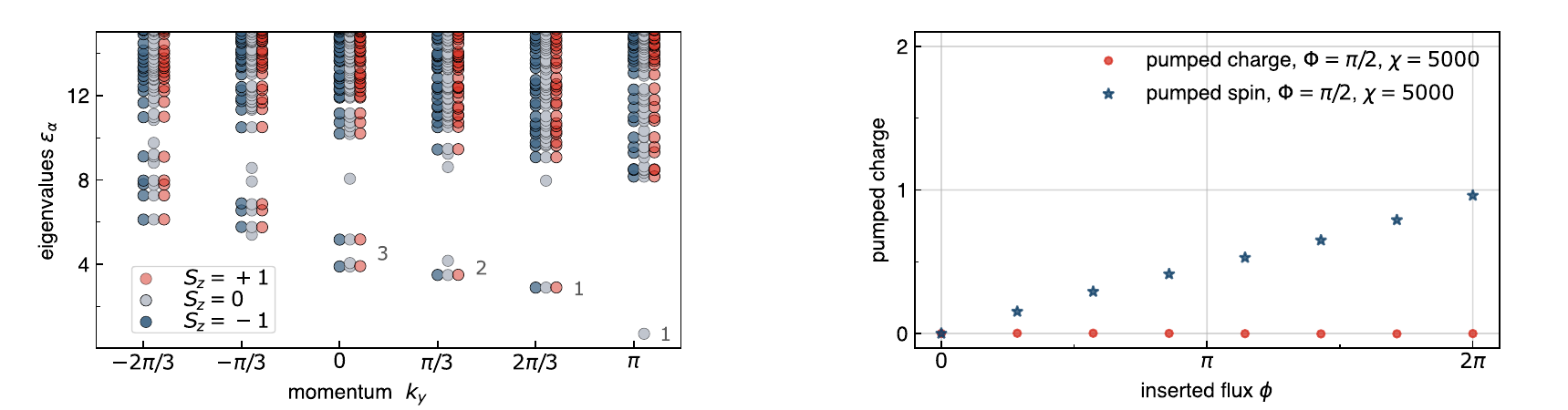}
\caption{\textbf{Effect of nearest-neighbor interactions on the entanglement spectrum and charge pumping.} For both panels $\Phi=\pi/2$, $U/V=6$, and $U=17t$. Left panel: Entanglement spectrum for $\chi=10000$. The symmetry-resolved spectrum and its degeneracies agree with the CSL prediction even for sizable nearest-neighbor interactions. Right panel: The system pumps an integer pseudo spin under flux insertion, confirming the topological nature of the spinon bands.}
\label{fig:Sq}
\end{figure*}
Nearest-neighbor interactions $V$, which are naturally present in 2D materials due to strong Coulomb repulsion, may destabilize the CSL. We now study the robustness of the observed CSL phases toward longer range interactions. To this end, we include repulsive longer-ranged interaction terms in the Hamiltonian, which then reads:
\begin{equation}
\begin{aligned}
\hat{H} &= - t \sum_{\mathclap{\langle ij\rangle, \sigma = \lbrace T,B \rbrace}} (e^{i \phi_{ij}} \; c^\dagger_{i,\sigma} c_{j,\sigma} + \mathrm{h.c.} )+ U \sum_i n_{i,T} \; n_{i,B} \\
&+ V\sum_{\langle ij\rangle} (n_{i,T} n_{j,T} + n_{i,B} n_{j,B} + n_{i,T} n_{j,B} + n_{i,B} n_{j,T}).
\label{eq:NNint}
\end{aligned}
\end{equation}
In the Mott insulating limit $U,V \gg t$, the nearest neighbor interactions indeed reduce the effective Heisenberg coupling in second-order degenerate perturbation theory
\begin{equation}
    J_H =4t^2/U \rightarrow  4t^2/(U-V),
\end{equation}
which points to a destabilization of the CSL. To study the robustness of the spin-liquid phase, we determine the ground state of Eq.~\ref{eq:NNint} using iDMRG, with a conservative estimate of $U/V =6$ and $U/t=17$, see App.~\ref{App:A}.   Having assumed a constant value for $U/V$, we significantly overestimate $V$ for smaller twist angles (i.e., larger moir\'e lattice constants), where the Hubbard model with local interactions only is an excellent approximation.
The topological features of the CSL remain for $U/V\geq 6$, as shown in Fig.~\ref{fig:Sq}. Our analysis therefore reveals a favorable experimental regime for the observation of spin liquid states at intermediate twist angles, as highlighted in the bottom panel of Fig.~\ref{fig:HubbardParams} in App.~\ref{App:A}. We note that the CSL regime could be enhanced even further by engineering the dielectric environment to maximize electronic screening as commonly assumed in the literature~\cite{Vishwanath2021,Kim2022}.

\textbf{Layer SU(2) symmetry breaking.---}
\begin{figure*}[t]
\includegraphics[width=0.95\textwidth]{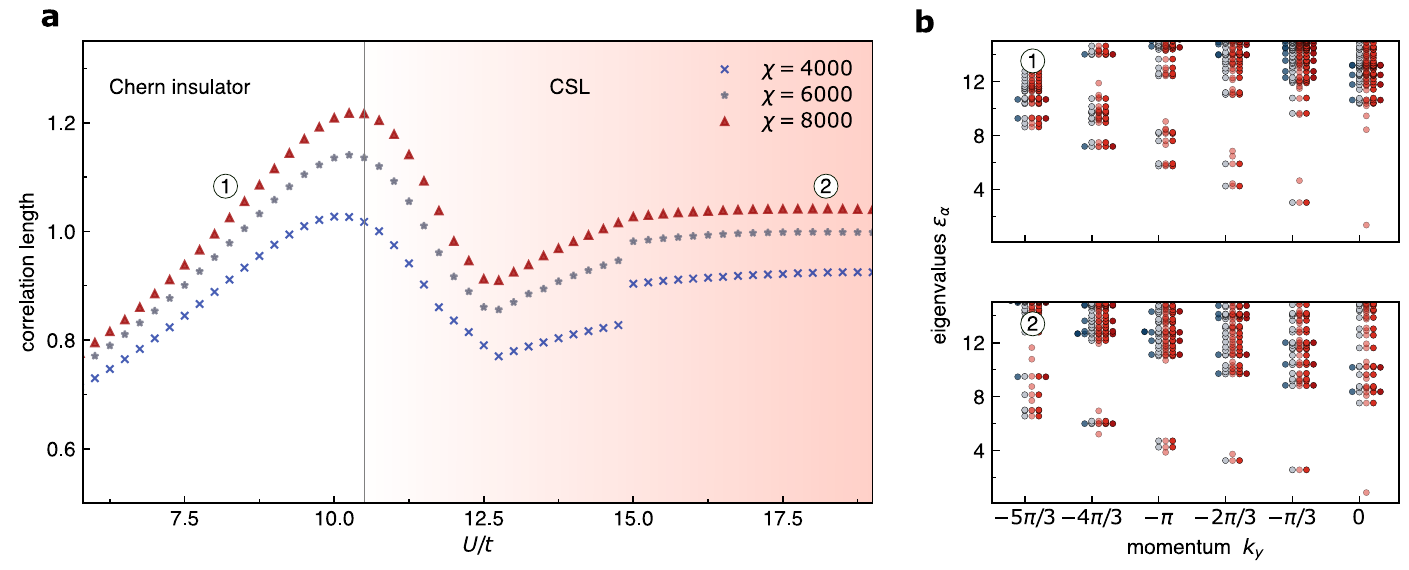}
\caption{\textbf{Triangular lattice Hubbard model with broken $SU(2)$ symmetry.} Ground state properties of the triangular lattice Hubbard model for $\delta = 0.1$ at $\Phi=\pi/2$ flux on an infinite cylinder with $L_y=6$. \textbf{(a)} Ground-state correlation length in the charge sector $(S_z,Q)=(0,0)$. As for the $SU(2)$-symmetric model, the correlation length is enhanced around $U/t \simeq 10.5$. \textbf{(b)}. Panel 1 and 2 show the ground state entanglement spectrum for $\chi=10000$ and charge $Q=0$ at small and large $U/t$, respectively. We find the characteristic eigenstate countings of a $\nu=2$ Chern insulator and the CSL, respectively. The degenerate $SU(2)$ multiplets remain well-defined on the edge, although the underlying Hubbard model is no longer symmetric.}
\label{fig:SU2_breaking}
\end{figure*}
While symmetries are not essential ingredients for the CSL, the extent to which $SU(2)$ can be broken before the spin liquid is destabilized is not clear a priori. Although it can be electrostatically ensured that $\sum_i\langle S^z(i)\rangle =0$, strain can induce hopping of different strengths in the two layers. The minimal model to study these perturbations is given by the following Hamiltonian:
\begin{equation}
\begin{aligned}
    \hat{H} = - &\sum_{\langle ij\rangle} (e^{i \phi_{ij}} \; t_T c^\dagger_{i,T} c_{j,T} + e^{i \phi_{ij}} t_B  c^\dagger_{i,B} c_{j,B} +\mathrm{h.c.})\\
    &+ U\sum_{i}n_{i,T} n_{i,B},
\end{aligned}
\end{equation}
where $t_{T} = t + \delta$ ($t_{B} = t - \delta$) label the tunneling strengths in the top (bottom) layer. The IQH state is expected to be robust as the wavefunction is close to a direct product of the two layers. For the CSL, the situation is less clear, as the two layers are correlated due to interactions. We find the leading order correction to the effective spin model describing the CSL by performing a Schrieffer-Wolff transformation, which yields the effective spin Hamiltonian:
\begin{equation}
\begin{aligned}
    \hat{H}_\text{eff}= &4\frac{t^2 + \delta^2}{U}\sum_{\langle ij\rangle} \mathbf{S}_i \cdot \mathbf{S}_j + 8 \frac{\delta^2}{U} \sum_{\langle ij\rangle} S^z_i S^z_j \\
    &+ \mathcal{O}(\frac{t^3}{U^2},\frac{\delta^3}{U^2}),
\end{aligned}
\end{equation}
where chemical potential terms of the form $\sim t \frac{\delta}{U} ( n_T - n_B)$ have already been compensated by external electric fields. $SU(2)$ breaking terms only appear at order $\mathcal{O}(\delta^2/U)$ which introduce some anisotropy. 
This is in accordance with the result of our DMRG simulations for $\delta = 0.1t$ and a flux of $\Phi=\pi/2$, for which we find very similar phase boundaries, as shown in Fig.~\ref{fig:SU2_breaking} a). The entanglement spectra shown in Fig.~\ref{fig:SU2_breaking} b),  show well-defined chiral edge modes that match the expected counting rules of the Chern insulator and the CSL, respectively. Due to the topological protection of the quantized Hall response, the edge theory remains well-defined and retains its $SU(2)$ multiplets even for sizable perturbations $\delta$.

\textbf{Ferromagnetic intersite exchange interactions.---}
\begin{figure*}[t]
\includegraphics[width=0.95\textwidth]{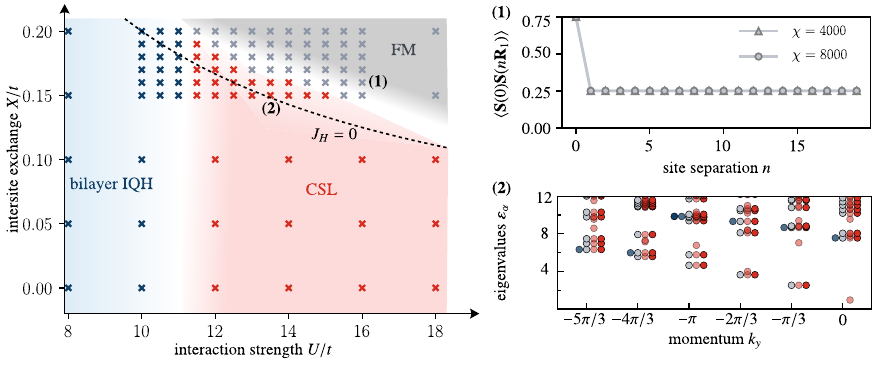}
\caption{\textbf{Phase diagram with intersite-exchange $X$ at $\Phi=\pi/2$ flux.} The ground state phase diagram as a function of the intersite exchange $X$ and onsite interaction $U$ of the triangular lattice Hubbard model reveals a ferromagnetic phase for large values of $X$ and $U$. At vanishing spin exchange $J_H$=0 (dashed line), the chiral spin liquid (CSL) phase is prevailing. Marked points are iDMRG data for $L_y=6$ and maximal bond dimension $\chi=8000$. For $U/t=16$ and $X/t=0.16$ \textbf{(1)} spin-spin correlations along the lattice vector direction $\mathbf{R}_1$ saturate at a finite positive value, indicating long-range ferromagnetic order. At $U/t=13.5$ and $X/t=0.15$ \textbf{(2)}, i.e., $J_H/t \approx 0$, the entanglement spectrum  shows the characteristic counting of the CSL.}
\label{fig:X}
\end{figure*}
Additionally to the previously discussed nearest-neighbor interactions $V$ and SU(2) symmetry breaking, we also explore the robustness of the CSL under nearest-neighbor intersite exchange interactions $X$:
\begin{equation}
    \hat{H} \rightarrow \hat{H} + X \sum_{\substack{\langle ij\rangle\\ \sigma, \sigma' \in \lbrace T,B \rbrace}} c^\dagger_{i,\sigma} c^\dagger_{j,\sigma'} c_{i,\sigma'} c_{j, \sigma}.
\end{equation}
The exchange term was argued to be relevant in typical moir\'e heterostructures and favors ferromagnetic spin exchange~\cite{Morales2022, Hu2021}. This is reflected in the direct contribution to the effective Heisenberg coupling 
\begin{equation}
    J_H = 4t^2/U - 2X,
\end{equation}
where $X>0$ for repulsive Coulomb interactions. Hence, for an estimate similar to the one discussed in Appendix~\ref{App:A} and with a conservative choice of parameters, one usually finds a ferromagnetic spin exchange $J_H<0$. In stark contrast, experimental measurements of the magnetization at half filling are consistent either with weak antiferromagnetic order~\cite{Tang2023a, Tang2023b} or no magnetic order~\cite{Ciorciaro2024} consistent with $J_H/t \gtrsim 0$. In Fig.~\ref{fig:X}, we map out the $X-U$ phase diagram for a flux of $\Phi=\pi/2$. We find that for large values of $U$, tuning the exchange $X$ leads to a transition from the CSL into a ferromagnetic phase as expected from the second-order expansion of $J_H$. Remarkably, in the experimental regime where $J_H/t \gtrsim 0$ (lower left part of the phase diagram), we find a robust CSL phase with clear signatures in the entanglement spectrum. We note that at this point, the spin liquid is not stabilized by the frustration of nearest-neighbor antiferromagnetic spin exchange on a triangular lattice but by the strong magnetic field that explicitly breaks time-reversal symmetry.

\section{Signatures of layer pseudo spin \label{sec:signatures}} 
Potentially the biggest challenge in spin-liquid physics is to find experimental signatures for their existence. Many of the previously proposed detection schemes of electronic-spin
liquid states---such as a quantum thermal Hall effect---are readily generalized to layer-spin systems. However, layer pseudo spin is easier to manipulate and probe which presents a crucial advantage and opens new avenues to detect the spin liquid: (I) By turning on an external magnetic field $B_z$ the ground state changes from a 120 degree N\'eel state at $B_z=0$ to the CSL, as sketched in Fig.~\ref{fig:1} c). Associated experimental signatures are the temperature dependence of the pseudo spin susceptibility $\chi(T)$. At high temperatures, it follows the net-Weiss law $\propto (T-\theta_{CW})^{-1}$, with negative Curie-Weiss temperature $\theta_{CW}$ due to antiferromagnetic interactions. At low temperatures, the N\'eel phase has a broad peak and approaches a constant value towards $T=0$. By contrast, the CSL has a spin gap $\Delta_S$, which leads to exponential decay of $\chi(T)$ below $k_B \Delta_S$. As we discuss below, $\chi(T)$ can be measured optically, which makes these signatures accessible in present-day experiments. (II) In stark contrast to intrinsic electronic spin, pseudo-spin can be electrically addressed, which allows one to couple the two pseudo-spin states to different electric fields (cf. discussion surrounding Fig.~\ref{fig:3} c) and d)). This enables counterflow measurements, which will exhibit a fractionally quantized Hall response, a hallmark of a CSL. While sample inhomogeneity and the difficulty in electrical isolation of the two layers render such a measurement rather demanding, it would be able to uniquely identify the topological order of the chiral layer-pseudo-spin liquid and distinguish it from all other phases in the phase diagram~\cite{Vishwanath2021}. (III) Topological properties of the CSL are encoded in the gapless modes on its edge~\cite{Wen1990}. In addition to spin-modes, one can measure the decay of electron correlation functions $\langle c^\dagger_{\sigma}(x,t) c_{\sigma'}(0,0)\rangle$ on the edge, which serve as additional probes of the spinons.

While observing fractional pseudo-spin Hall conductivity requires transport experiments, other signatures of the CSL could be obtained using all-optical measurements, which overcome several obstacles through tight focusing of the probe lasers into regions where the moir\'e potential is uniform. The spin-susceptibility at finite field $E_z$ can be extracted via the attractive polaron resonance strength of each layer, which measures how many electrons the exciton can form a bound trion state with and thereby directly depends on the electron density~\cite{sidler_polaron,Smolenski2019}. Probing the pseudo-spin magnetization $\langle S_z \rangle$ is then possible since each layer generically has a different exciton resonance\footnote{This is true also for twisted homobilayer structures where strain lifts the degeneracy of the exciton resonance.}. As outlined above, measurement of magnetization as a function of $E_z$ at low temperatures $k_B T \ll \Delta_s$ should show the existence of a spin-gap. The appearance of the charge gap in turn, is evidenced by the cusps in attractive and repulsive polaron resonances, indicating the modification of dynamical screening of excitons by electrons~\cite{yuya_20}. Such local probes are particularly relevant for near term devices as they are readily accessible and largely insensitive to large-scale disorder. Further, direct evidence for the emergence of a gauge field could be verified using optical Hall measurements. It has been demonstrated that effective electric fields can be imprinted on excitons using crossed magnetic and time dependent electric fields~\cite{Lim2017}. A combination of such an effective dipole electric field and the emergent gauge field should result in layer-contrasting Hall effect~\cite{Yao2022}, which can be measured by determining the spatial dependence of the attractive polaron resonance in each layer along the axis that is orthogonal to the dipole electric field. Furthermore, novel momentum-resolved techniques unique to 2D materials could also allow for the measurement of ARPES-like spectra~\cite{Ilani2022} which can provide signatures of spinon excitations in quantum spin liquids~\cite{Kadow2022}.

Complementary insights could be obtained by measuring $\langle S^z_{\mathbf{q}=0}(t) S^z_{\mathbf{q}=0}(0) \rangle$ using correlations between resonantly scattered photons on the attractive polaron resonance. If coherent optical Raman manipulation of layer pseudo-spin is possible, then suppression of spin noise along all axes for $k_B T \ll \Delta_s$ can be measured~\cite{priv-comm-gomez}.

\section{Conclusions and Outlook \label{sec:outlook}}
We have shown how a large class of quantum phases and transitions can be studied in multi-layer TMDs. In particular, topologically ordered CSLs can be stabilized by magnetic fields utilizing the layer degree of freedom as a synthetic spin~\cite{Vishwanath2021,schwartz21,Kuhlenkamp21}. The absence of a magnetic Zeeman effect for the pseudo-spin allows us to target topological states by controlling the strength of ring exchange processes using large external magnetic fields. For weak interactions a variety of Hofstadter states can be prepared by sweeping $B_z$, while a CSL forms for intermediate interaction strengths. At specific fluxes, our model realizes topological insulator-insulator Mott transitions. Understanding the details of these transitions is an interesting direction for future work and will help to better understand the phase boundaries of the CSL. Remarkably, the field induced CSL is found to be exceptionally robust and occupies a large region of the phase diagram. Combined with the electric tunability and layer-selective read-out of layer pseudo-spin, this makes TMD heterostructures particularly promising platforms to study spin liquid physics. Novel probing schemes unique to the pseudo-spin degree of freedom offer an additional advantage of these systems over conventional spin liquid candidates. Competing spin and charge ordered phases can be more easily identified; while counterflow measurements~\cite{Vishwanath2021} directly probe the topology of the spin liquid. It is also possible to find fingerprints of the spin liquid phase with all optical measurements, which provide local probes that are crucial for near-term devices.

Our results open up several theoretical and experimental avenues to study topological order and exotic phase transitions. For one, Mott insulators can be stabilized at fractional fillings by longer-ranged interactions, leading to spin systems with different lattice geometries and more exotic states for small interaction strengths. More specifically, for $\Phi=\pi/3$ densities of $n_e=1/3, 1/9, \dots$ realize excitonic insulator candidates and FQH states respectively, both of which eventually transition to frustrated spin states in Mott-Wigner insulators at large $U/t$. 
This poses interesting questions about the nature of quantum phase transitions between topological order in the charge and spin sectors.
Furthermore, in the weak field limit a variety of interesting competing states emerges, and their interplay is largely unexplored. Most notably, this regime is expected to feature Hofstadter physics, quantum Hall ferromagnetism and $SU(4)$ spin liquids~\cite{Vishwanath2021}.

\appendix
\begin{figure}[b]
\includegraphics[width=0.98\columnwidth]{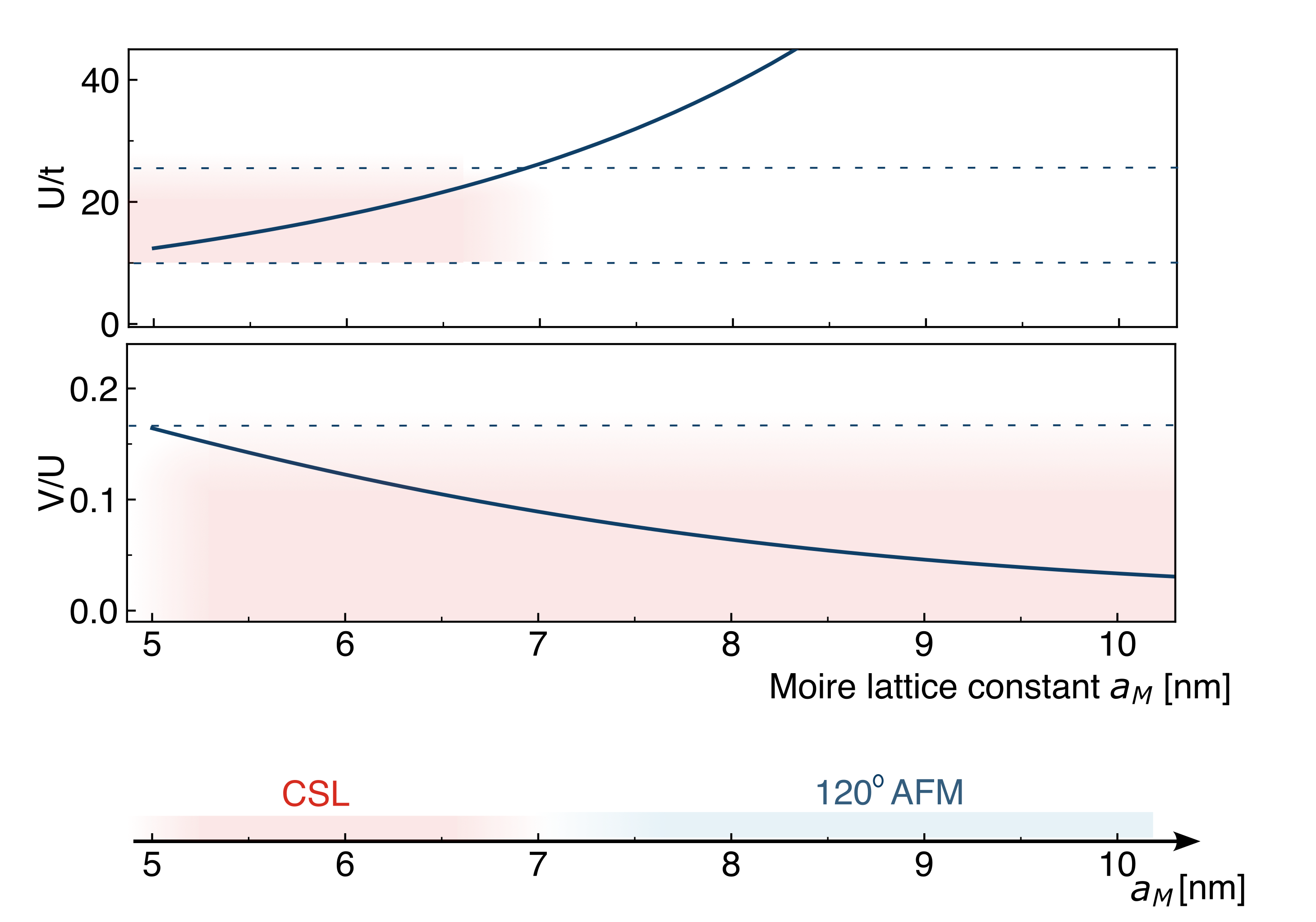}
\caption{\textbf{Hubbard model parameters.} $U/t$ and $V/U$ for a WSe$_2$ system as a function of lattice constant $a_M$, assuming $d=2$nm, $\psi = -94^\circ$ and $\epsilon=7$. Favorable regimes for the CSL are shaded in red. \textbf{Top:} Increasing $a_M$ drives the system deeper into the Mott phase which eventually gives rise to a 120 degree magnet once $J_H$ dominates (upper dashed line), while the system tends to form an integer quantum Hall state for $U/t \leq 10$ (lower dashed line), which we determine from our numerical simulations. \textbf{Middle:} Nearest-neighbor interactions are strongly suppressed for large $a_M$. We find numerically that the CSL remains robust when $V/U\leq 1/6$ as indicated by the dashed line. \textbf{Bottom:} Resulting phase diagram as a function of $a_M$, indicating the regime for which both on-site and nearest-neighbor interactions are favorable for the CSL as determined from the $t-U-V$ Hubbard model. }
\label{fig:HubbardParams}
\end{figure}

\section{Determining the effective Hubbard model}\label{App:A}

We derive the parameters of the Hubbard model from microscopic properties of the TMDs, following the procedure outlined in Ref.~\cite{Wu18}. In the set-up discussed in the main text, holes in the two doped TMD layers feel a potential energy variation $V_M(\mathbf{r})$ which is well approximated by the lowest few harmonics $\lbrace \mathbf{g}_j | j = 1,\dots,6\rbrace $ in the limit of small twist angles
\begin{equation}
    V_M(\mathbf{r}) = \sum_{j=1}^6 v_{\mathbf{g}_j} e^{i \mathbf{g}_j \cdot \mathbf{r}},
\end{equation}
where the lowest harmonics of the potential are specified by $ v_{\mathbf{g}_1}= v_0 e^{i\psi}$.  In the low-doping limit, we approximate the valence band dispersion of the TMDs as parabolic and solve the corresponding Schrödinger equation to determine the moir\'e bands and Bloch functions $u_n(\mathbf{q})$. The hopping parameter $t$ in the Hubbard model is determined by fitting a tight-binding dispersion to the lowest-lying Moire band. We calculate the interaction parameter $U$ by employing a projective construction~\cite{Vanderbilt12,Zhang19} to determine a complete set of localized Wannier functions for the lowest band $w_{\mathbf{R}_n}(\mathbf{r})$, where $\mathbf{R}_n$ labels the position of the unit cell. This yields
\begin{equation}
    U = \int d^2\mathbf{x}\int d^2 \mathbf{y} \; |w_{\mathbf{R}_n}(\mathbf{x})|^2\; V_C(\mathbf{x}-\mathbf{y}) \;|w_{\mathbf{R}_n}(\mathbf{y})|^2,
\end{equation}
and a similar expression for the strength of nearest-neighbor interactions
\begin{equation}
    V = \int d^2\mathbf{x}\int d^2 \mathbf{y} \; |w_{\mathbf{R}_n}(\mathbf{x})|^2\; V_C(\mathbf{x}-\mathbf{y}) \;|w_{\mathbf{R}_{n+1}}(\mathbf{y})|^2.
\end{equation}

Where we assume an electrostatic Coulomb potential
\begin{equation}
   V_C(\mathbf{r}) = \frac{e^2}{4\pi \epsilon_0 \epsilon} \left(\frac{1}{|\mathbf{r}|} - \frac{1}{\sqrt{\mathbf{r}^2 + 4d^2}} \right),
\end{equation}
which takes into account the dielectric constant of the environment $\epsilon$ and that electrons induce mirror charges in a metallic gate separated by a distance $d$, which screens the Coulomb law down to dipolar interactions $V_D(\mathbf{r \gg d})\approx 2 e^2 d^2 / 4\pi \epsilon_0\epsilon |\mathbf{r}|^3$ when their separation is much larger than the distance to the gate. Longer-range interaction terms beyond $U$ and $V$ can be neglected as they are strongly suppressed for the parameters we consider. For concreteness, we focus on a trilayer WSe$_2$/MoSe$_2$/WSe$_2$ setup in the following and assume a hole mass of $m^* = 0.35$~\cite{Kormanyos15}, a potential strength $v_0 \simeq 10$meV and $\psi = -94^\circ$ as suggested by DFT calculations~\cite{Wu18}, although our results do not depend strongly on the precise values of $v_0$ and $\psi$, and apply qualitatively also to the other setups discussed in the main text. We consider a dielectric constant of $\epsilon = 7$, which is relatively small and thus realistic for the encapsulation with hBN.
The functional dependence of $U/t$ and $V/U$ on the Moire lattice constant $a_M$ is shown in Fig.~\ref{fig:HubbardParams}, for $d=2$nm. For $U/t \gtrsim 25$ ($a_M\gtrsim 7$nm) the Heisenberg coupling dominates and the system forms a 120 degree state~\cite{Wietek17}. In the limit of small lattice constants, or larger twist angles, the system eventually forms a bilayer IQH state. For intermediate lattice constants our numerical analysis suggests the CSL is stable over a sizable regime even in the $t-U-V$ Hubbard model. The robustness of the CSL to perturbations is studied in detail in Sec.~\ref{Sec:robust} of the main text.

\vspace{.5cm}
\textit{\textbf{Acknowledgments.---}} We thank M. Drescher, K. Fai Mak, J. Feldmeier, J. Hauschild, F. Pollmann, J. Shan, T. Smolenski, and A. Young for fruitful discussions. 
We acknowledge support from the Deutsche Forschungsgemeinschaft (DFG, German Research Foundation) under Germany’s Excellence Strategy-EXC-2111-390814868 and DFG grant No. KN1254/2-1, No. KN1254/1-2, the European Research Council (ERC) under the European Union’s Horizon 2020 research and innovation programme (grant agreement No. 851161), as well as the Munich Quantum Valley, which is supported by the Bavarian state government with funds from the Hightech Agenda Bayern Plus. The work at ETH Zurich was supported by the Swiss National Science Foundation (SNSF) under Grant Number 200021-204076. C.K. was supported by an ETH Zurich Doc.Mobility Fellowship.

\textbf{\textit{Data and materials availability.---}} Data analysis and simulation codes are available on Zenodo upon reasonable request~\cite{zenodo}.

%

\clearpage

\onecolumngrid
\begin{center}
\textbf{\Large{\large{Supplemental Material: \\Chiral Pseudo Spin Liquids in Tunable Moir\'e Heterostructutes}}}
\end{center}
\maketitle
\setcounter{figure}{0}
\setcounter{equation}{0}
\renewcommand{\thepage}{S\arabic{page}} 
\renewcommand{\thesection}{S\arabic{section}} 
\renewcommand{\thetable}{S\arabic{table}}  
\renewcommand{\thefigure}{S\arabic{figure}} 
\renewcommand{\theequation}{S\arabic{equation}} 
\onecolumngrid
\vspace{-1cm}

\subsection{Gauge choice for different flux values}
\begin{figure}[h]
\includegraphics[width=0.86\columnwidth]{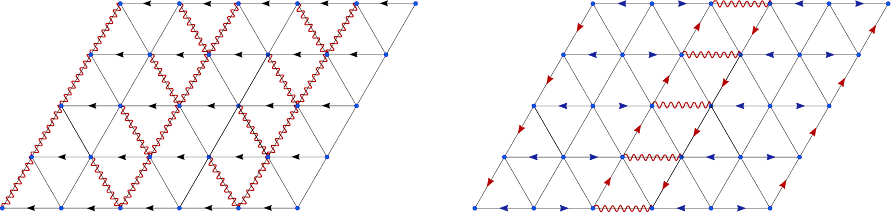}
\caption{\textbf{Gauge choices for different fluxes.} Flux patterns on the triangular lattice leading to $\Phi=\pi/2$ \textbf{(a)} and $\Phi=\pi/3$ \textbf{(b)}. Hopping along a bond in the direction of a red (blue) arrow yields a phase factor of $e^{i2\pi/3}$ ($e^{i\pi/3}$), while hopping along a black arrow yields a phase of $e^{i\pi/2}$. Squiggly bonds carry a factor of $-1$.}
\label{fig:S1}
\end{figure}

To implement a static magnetic field on the triangular lattice we perform a Peierls substitution $t_{ij}\rightarrow t_{ij} e^{i\phi_{ij}}$, where the choice of phases $\phi_{ij}$ is not unique. To accommodate for the $\Phi=\pi/2,\pi/3$ flux per triangle and to keep translational invariance in $y$ direction, we perform our simulations on unit cells of size $L_x=2,3$. In the main text, we chose gauges leading to the phase patterns shown in Fig.~\ref{fig:S1}. As any gauge choice breaks explicit translational invariance, we work with unit cells and cylinder geometries which are commensurate with the flux. 

\subsection{Ground state crossing for $\Phi=\pi/3$}
\begin{figure}
\includegraphics[width=0.7\columnwidth]{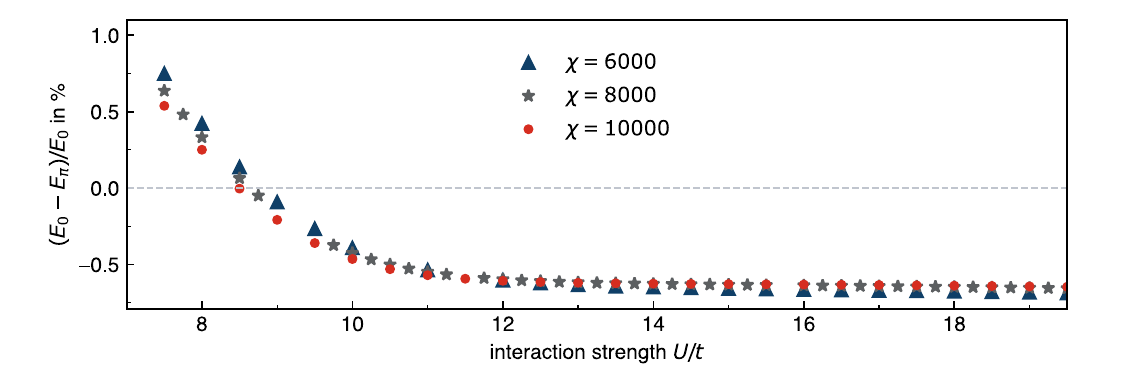}
\caption{\textbf{Energy difference of the lowest lying $k_y=0$ and $k_y=\pi$ states for $\Phi=\pi/3$.} We find that for $U\simeq 8.5t$ the ground-state energies of the two momentum sectors cross. This point moves to weaker interactions as the bond dimension increases. This level crossing occurs in the vicinity of the enhanced correlation length observed in the main text. We expect that in the CSL phase both eigenstates are degenerate in the thermodynamic limit.}
\label{fig:S2}
\end{figure}

We can reliably prepare states in the $k_y=\pi$ and $k_y=0$ sectors on the $L_x=3$ unit cell and flux $\Phi = \pi/3$. As discussed in the main text, the two ground states of the CSL sectors lie at $k_y = 0,\pi$ but are slightly split due to finite size effects. Therefore, in our truncated system the transition from the Hall states to the CSL occurs in two steps: First the correlation length in the $k_y=\pi$ sector diverges and shortly thereafter its energy crosses with the $k_y=0$ state and becomes the true finite-size ground state. We show the difference in ground state energies as a function of interaction strength and bond dimension in Fig.~\ref{fig:S2}. We find that the crossing takes places around $U\sim 8.75t$, where the correlation length of the $k_y=\pi$ state is still very large, as discussed in the main text.
\subsection{Characterizing the phases for $\Phi=\pi/3$}
While the precise value of the flux is important in the Hofstadter regime, as it determines the filling of the Chern bands, in the Mott insulating limit it simply controls the strength of the chiral spin term $J_\chi$. Therefore, we generically expect chiral spin liquids to appear in the Mott insulator, as long as $J_\chi~\sim \sin(\Phi)$ is large enough. Here we discuss the fate of the Mott insulator for $\Phi=\pi/3$, which corresponds to a filling of $\nu_T=3$.
\begin{figure}
\includegraphics[width=1\textwidth]{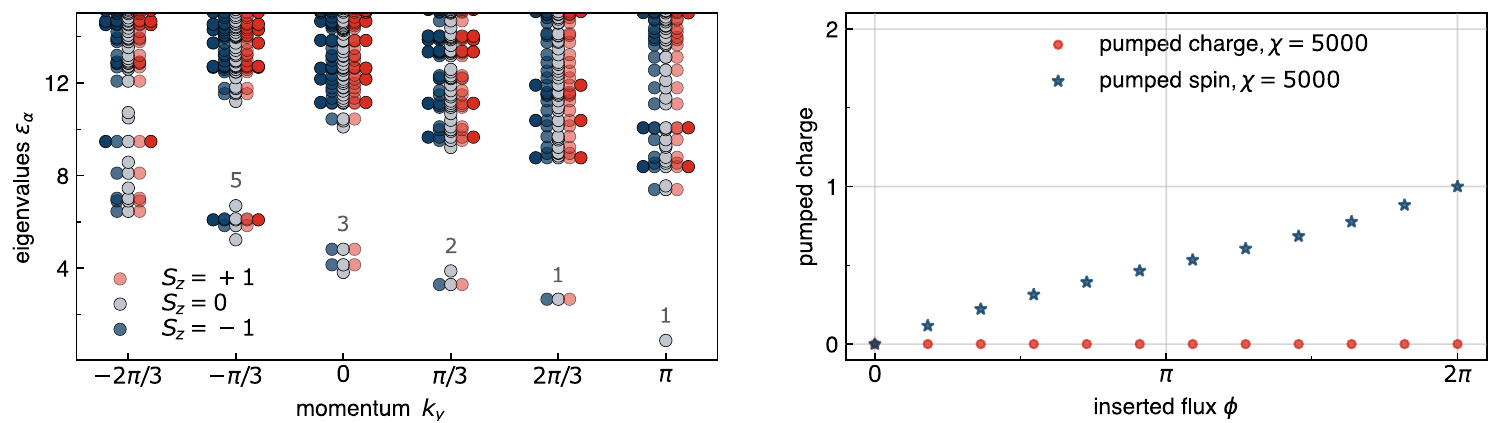}
\caption{\textbf{Entanglement spectrum and charge pumping for $\Phi=\pi/3$ and $U=15t$.} Left panel: Entanglement spectrum for $\chi=10000$. The symmetry resolved spectrum and its degeneracies agree with the $SU(2)_1$ WZW edge theory of the CSL. Right panel: The system pumps an integer pseudo spin under flux insertion, confirming the topological nature of the spinon bands.
}
\label{fig:S3}
\end{figure}
Following the analysis done for $\Phi=\pi/2$ in the main text, we compute the entanglement spectrum and analyze the pumped charges under flux insertion. Our results in the Mott limit are shown in Fig.~\ref{fig:S3} for $U=15t$. The entanglement spectrum agrees with the spectrum of the $SU(2)_1$ WZW model describing the spin-density-wave excitations on the edge (left panel). The absence of gapless density-wave excitations, in addition to the vanishing pumped charge under the insertion of real magnetic flux (right panel) indicates that charges are frozen and hence that the state is insulating. Under the insertion of opposite fluxes in the two layers, i.e., when coupling to charge-neutral, spin excitations, we find a fractional spin Hall effect, as discussed in the main text~\cite{Laughlin87,Moore2020}. As expected, no charge is pumped under the insertion of opposite fluxes, and no spin is pumped under symmetric flux insertion (not shown). Hence, the state describes a CSL. These results demonstrate that the CSL is a generic feature of the Hofstadter-Hubbard model at strong coupling independent of the specific values of the magnetic field.

\begin{figure}
\includegraphics[width=\columnwidth]{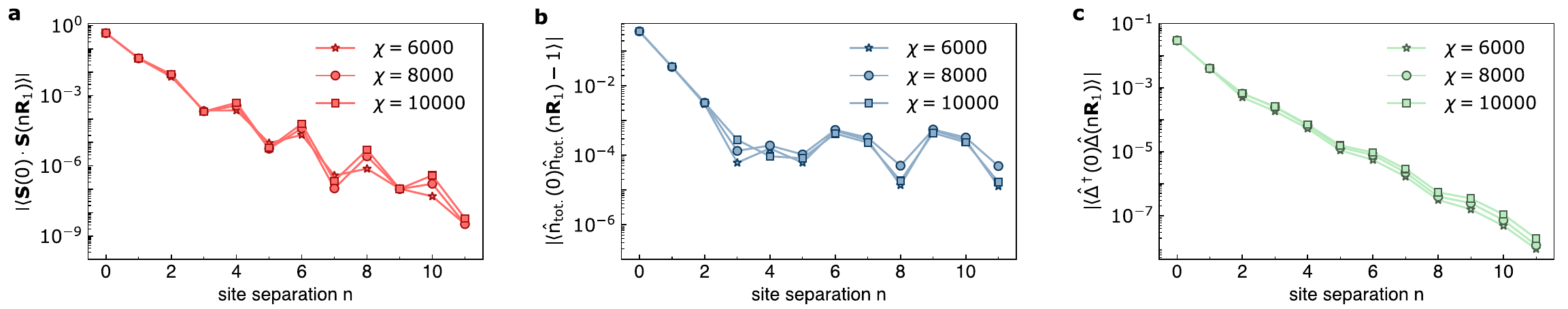}
\caption{\textbf{Correlations in the weakly interacting phase for $\Phi=\pi/3$.} Correlation functions are plotted for $U/t=3$ as a function of distance along the x-direction. \textbf{a)} Spin-correlations are exponentially decaying, showing the absence of long-range spin order. \textbf{b)} Density correlations show an initial exponential decay and a weak residual tendency to form charge order. \textbf{c)} Superconducting correlations are small and decay exponentially. This indicates that the state may be a featureless Hall state.}
\label{fig:S4}
\end{figure}

The fate of the Hofstadter states in the presence of weak interactions is a priori not obvious. For $\Phi=\pi/3$, we find that the system no longer exhibits quantized charge or spin pumping in the charge-liquid regime. To study the propensity of the state to develop long range order, we analyze spin, density and superconducting correlation functions, which are shown in Fig.~\ref{fig:S4}.
The formation of excitonic insulators and other layer-spin ordered states is ruled out, as we find exponentially decaying spin correlations, see Fig.~\ref{fig:S4} a). Similarly, density correlations show an initial exponential decay before saturating at a weak residual value of $\approx 10^{-4}$, which indicates weak charge order, shown in Fig.~\ref{fig:S4} b). Since this value is very small and the circumference of the cylinder is finite, we attribute this to finite size effects.

In the absence of magnetic fields, superconductivity has been discussed as a possible phase in the triangular Hubbard model~\cite{Gannot2020}. The tendency to become superconducting is captured by pairing terms of the form:
$\hat{\Delta}(x,y) =(c^\dagger_{T,x,y+1}c^\dagger_{B,x,y} -c^\dagger_{B,x,y+1}c^\dagger_{T,x,y})/\sqrt{2}$ and the corresponding correlation functions are depicted in Fig.~\ref{fig:S4} c). This shows that correlations decay exponentially and we do not find signatures of superconducting order. We also did not observe superconducting instabilities in other correlation functions. Our analysis therefore suggests that the $\Phi=\pi/3$ state in the half-filled triangular lattice is consistent with a featureless Hall state.

\subsection{Probing the other sectors of the CSL}
\begin{figure}
\includegraphics[width=\columnwidth]{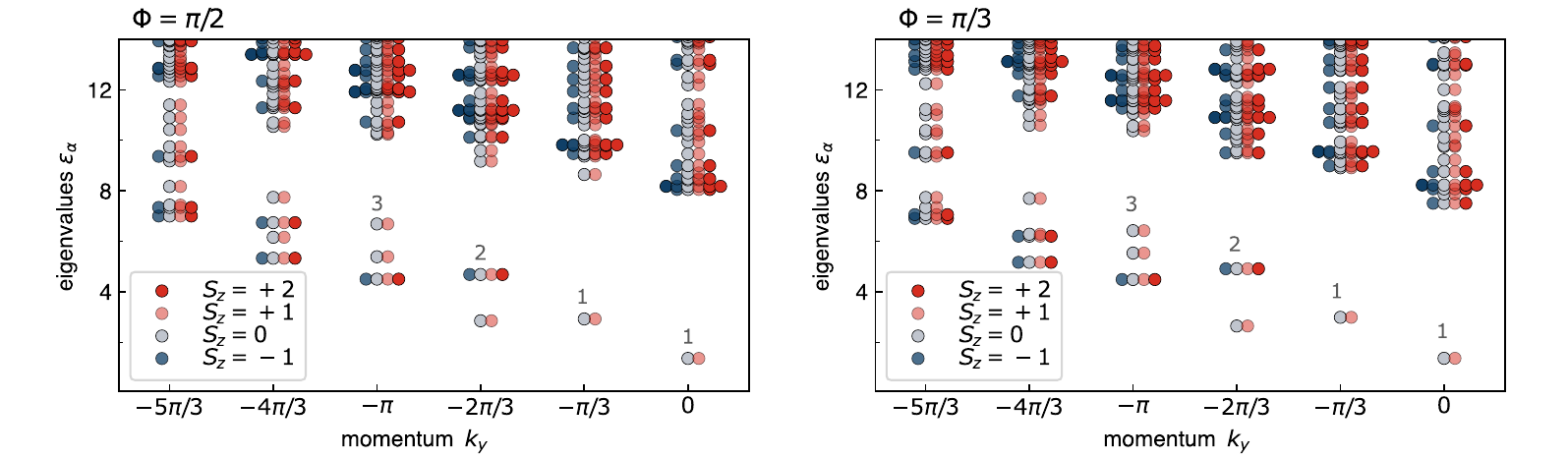}
\caption{\textbf{Entanglement spectra in the semion sector.} Entanglement spectra in the CSL phase for $U=15t$ and $\chi=10000$. For both $\Phi=\pi/2$ (left panel) and $\Phi=\pi/3$ (right panel) the entanglement spectra obey the (1,1,2,3,5,\dots) counting rule for the edge theory of the CSL. In contrast to the trivial sector, the lowest lying state is now part of a spin $1/2$ representation.}
\label{fig:S5}
\end{figure}

In the fermionic system, the two topological ground states of the CSL carry momentum $k_y=0$ and $k_y=\pi$. Due to finite size effects these states are not perfectly degenerate~\cite{Sheng2017,Moore2020}. We prepare these states in our iDMRG simulations by initializing them with product states carrying $\pi$ momentum in the y-direction. We show the entanglement spectra of the resulting ground states in Fig.~\ref{fig:S5}.
 As for the trivial sector, we observe a clear separation between bulk and edge modes and the number of dominant eigenvalues in each spin sector continues to follow the (1,1,2,3,\dots) counting rule. The lowest lying states of the $SU(2)_1$ algebra are now doubly degenerate and are part of a spin $1/2$ representation rather than a singlet, see Fig.~\ref{fig:S5}. We remark that instead of preparing finite momentum states via the initial conditions, the two sectors can also be interchanged by threading magnetic flux through the cylinder.

\end{document}